\documentclass{aa}
\usepackage{psfig}
\usepackage{graphicx}

\newcommand{\gtrsim}{\mathrel{\hbox{\rlap{\lower.55ex \hbox {$\sim$}}
                   \kern-.3em \raise.4ex \hbox{$>$}}}}
\newcommand{\lesssim}{\mathrel{\hbox{\rlap{\lower.55ex \hbox {$\sim$}}
                   \kern-.3em \raise.4ex \hbox{$<$}}}}

\begin{document}

   \title{A superburst from GX\,3+1}

   \author{Erik Kuulkers
           \inst{1,2}
          }


   \institute{
	      SRON National Institute for Space Research,
	      Sorbonnelaan 2, 3584 CA Utrecht, The Netherlands
	      \and
	      Astronomical Institute, Utrecht University,
	      P.O.\ Box 80000, 3508 TA Utrecht, The Netherlands
             }

   \date{Received --; accepted --}

   \titlerunning{A superburst from GX\,3+1}

\abstract{I found one long X-ray flare from the X-ray burster GX\,3+1 in almost 6~years of
observations with the RXTE All Sky Monitor (ASM).
The event had a peak flux of about 1.1~Crab (1.5-12\,keV), lasted between 
4.4 and 16.2~hours and exhibited a fluence 
of more than about 5$\times$10$^{41}$\,erg for a source distance of 5\,kpc. During the 
exponential-like decay, with an exponential decay time of 1.6~hours, spectral softening is seen.
The total ASM effective exposure time on GX\,3+1 is estimated to be around a year.
The flare bears all the characteristics of the recently discovered so-called superbursts in other
X-ray burst sources.
       \keywords{accretion, accretion disks --- binaries: close --- stars: individual (GX\,3+1) --- 
       stars: neutron --- X-rays: bursts}
  }

   \maketitle

\section{Introduction}

Recently, six long X-ray flares lasting several hours have been identified
in five low-mass X-ray binaries (Cornelisse et al.\ 2000, 2001,
Strohmayer \&\ Brown 2001; Wijnands 2001; Kuulkers et al.\ 2001).
For one source, 4U\,1636$-$53, two such events were reported which
occurred about 4.7~years after each other (Wijnands 2001). 
So far, these flares have only been seen in
X-ray bursters with persistent pre-flare luminosities of $\simeq$0.1--0.3 times
the Eddington luminosity (Wijnands 2001; Kuulkers et al.\ 2001).

The long X-ray flares share many of the characteristics of type~I X-ray
bursts\footnote{Type~I X-ray bursts have light curves with a 
rise which is faster than the exponential-like decay; their
emission is well described by black-body radiation with temperatures, $kT$, around
2\,keV and apparent black-body radii around 10\,km; they show X-ray spectral softening 
during the decay. They have durations of seconds to minutes. For a review, see Lewin et al.\ (1993).}
and are, therefore, attributed to thermonuclear runaway events on
a neutron star (e.g.\ Cornelisse et al.\ 2000). The differences
with type~I bursts are their long duration (exponential decay times of
a few hours), their large fluences (about 10$^{42}$\,erg), and their
rarity. Because of the large fluences, the
flares are referred to as `superbursts'.
A likely fuel for the superbursts is carbon, left over 
from stable and unstable hydrogen and/or helium burning 
(Cumming \&\ Bildsten 2001; Strohmayer \&\ Brown 2001). Unstable electron
capture by protons with subsequent capture of the resulting neutrons by heavy nuclei
is another conceivable option (Kuulkers et al.\ 2001).

\begin{figure}
\resizebox{\hsize}{!}{\includegraphics[clip, bb=38 254 421 717]{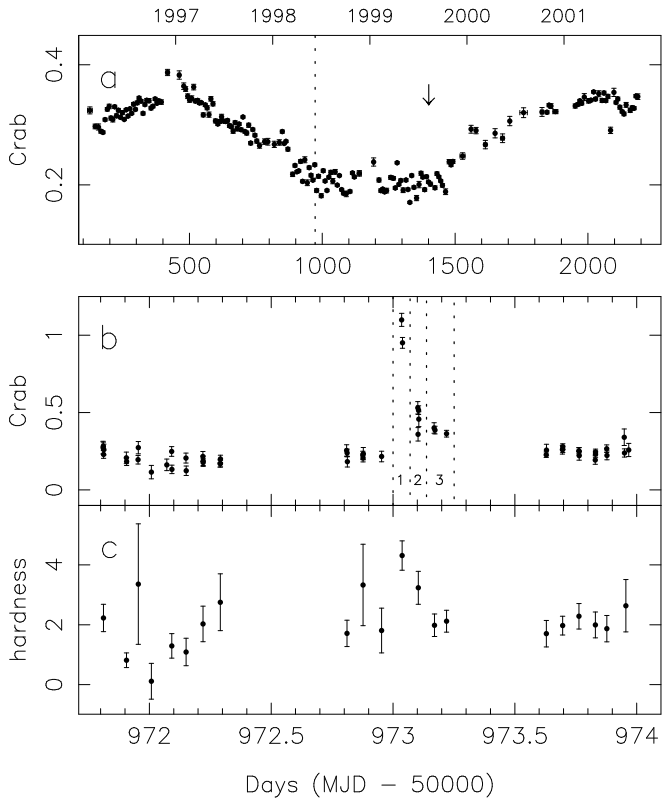}}
\caption{
{\bf a)} RXTE/ASM (1.5--12\,keV) light curve of GX\,3+1 from
1996 (MJD\,50088) to 2002 (MJD\,52206). The data shown represent the mean of 7 consecutive daily 
averages and are normalized to the Crab count rate (75\,cts\,s$^{-1}$\,SSC$^{-1}$). 
With a dotted line the time of the X-ray flare is shown. The time of the radius-expansion 
type~I X-ray burst observed with the RXTE/PCA (Kuulkers \&\ van der Klis 2000) is indicated 
by an arrow. 
{\bf b)} Light curve of the individual RXTE/ASM dwells (normalised to Crab units; 1.5--12\,keV) 
around the time of the flare.
The dotted lines mark the three intervals used in the spectral analysis, see text.
{\bf c)} Hardness curve around the time of the flare. Hardness is defined as the ratio
of the count rates in the 5--12\,keV and 1.5--3\,keV energy bands. Shown are averages 
of data which were less than 30~min apart, see text.}
\label{plot_asm}
\end{figure}

I here report on a superburst seen with the {\it Rossi X-ray Timing Explorer} (RXTE)
All Sky Monitor (ASM) from GX\,3+1 in June 1998.
For a preliminary announcement of this event see Kuulkers (2001).
The overall X-ray intensity of the GX 3+1 varies slowly on time scales of
months to years by a factor of about 2
(e.g.\ Makishima et al.\ 1983; see Fig.~\ref{plot_asm}a). 
Type~I X-ray bursts in GX\,3+1 were first discovered by Hakucho (Makishima et al.\ 1983). 
A type~I X-ray burst with radius expansion (due to the burst luminosity reaching the Eddington limit) 
was observed by the Proportional Counter Array (PCA) onboard RXTE, 
enabling one to estimate the distance to the source to be in the range
4--6\,kpc (Kuulkers \&\ van der Klis 2000).

\section{Observations and Analysis}

The ASM (Levine et al.\ 1996) 
is one of the three instruments onboard RXTE. It consists of three Scanning 
Shadow Cameras (SSCs) mounted on a rotating drive such that the center
of the field of view of one camera is perpendicular to that of the other two.
The cameras are held stationary for 90-s intervals, called
`dwells', during which data are accumulated.  The mount rotates the assembly
through a 6$\degr$ angle between dwells until a full rewind is necessary.  In
this manner about 80\%\ of the sky is observed every 90-min orbit around the earth.  
Each dwell provides intensities in the 1.5--12\,keV band for all known sources in the field
of view of each camera, and results that meet a set of reliability criteria are
saved in electronic tables that are available over the World Wide
Web.

The ASM data are also available in three sub-bands: 1.5--3\,keV, 3--5\,keV, and 5--12\,keV.
The boundaries are, however, crude approximations. The actual channel boundaries 
change somewhat with time, anode, software changes, etcetera. Moreover, the amount 
of these changes differ between the three SSCs 
(A.~Levine 2001, priv.\ comm.).
A pre-flight instrument effective area matrix for a source at the center of the field of
view of an SSC is available, providing the detector response between 
1 and 20\,keV on an energy grid with a 
resolution of 0.1\,keV\footnote{Note that this matrix is also used by {\sc PIMMS}
({\tt http://legacy.gsfc.nasa.gov/docs/software/tools/ pimms.html}).}.
It was used for estimating data system properties and rough sensitivities
during the ASM development stage, and was certainly not designed for doing X-ray 
spectral fits to actual observations in three energy bands. There were small changes in the 
instrument design after the matrix numbers were computed, but the numbers were never updated
(A.~Levine 2001, priv.\ comm.).
However, the matrix can be still useful for rough calculations, 
as done here. Note that a simple spectral model such as a black-body with only two free 
parameters may still be constrained by data in three energy bands.

The Crab (pulsar and nebula) is one of the main bright and steady X-ray calibration sources. 
Its observed ASM count rate is about 75\,cts \,s$^{-1}$\,SSC$^{-1}$ (1.5--12\,keV,
when at the center of an SSC field of view; see e.g.\ Levine et al.\ 1996). 
The pre-flight instrument effective area matrix,
however, predicts a Crab count rate of about 88\,cts \,s$^{-1}$\,SSC$^{-1}$ in the same energy 
band. (The Crab spectrum in the energy band of interest can be described by an 
absorbed power-law;  
for the parameter values see e.g.\ Schattenburg \&\ Canizares (1986), and 
references therein.)
For doing X-ray spectral analysis one, therefore, has to renormalize the source count rates.
The observed source count rates in a certain energy band were multiplied by the ratio of the 
predicted Crab count rate and the average observed Crab count rate in the same energy band. 
The latter were determined from the individual dwells during an interval 
centered around the observations, i.e\ MJD\,50900--51050 
(1998 March 28 to August 25). The resulting source count rates were then input for 
my spectral fitting. 
The rms in the average Crab count rates was included in the uncertainties
of the source count rates (4.5\%, 13\%, 5\%\ and 5\%, in the 1.5--12\,keV, 1.5--3\,keV,
3--5\,keV and 5--12\,keV bands, respectively). Note that 
the errors in the individual dwell measurements contain already, apart from
counting statistics, a 3\%\ systematic error; according to the ASM team
this will likely be an underestimate of the error in many cases.
The method outlayed above is in principle only applicable if the source spectrum is 
comparable to that of the Crab. However, since I am only interested in rough estimates 
my 
procedure is 
adequate.

In the spectral analysis 
the interstellar absorption column, N$_{\rm H}$,
was fixed at 1.7$\times$10$^{21}$\,cm$^{-2}$ (Christian \&\ Swank 1997).
Whenever the goodness
of fits in the spectral or temporal analyses, expressed as reduced $\chi^2$ ($\chi^2_{\rm red}$) 
for certain degrees of freedom (dof), exceeded 1, the error bars in the 
data points were arbitrarily increased so that $\chi^2_{\rm red}=1$. 
I then determined uncertainties in the fitted
parameters using $\Delta\chi^2=1$.

\section{Results}

By visually inspecting the ASM 1.5--12\,keV database of GX\,3+1 a single flare was found
on 1998 June 9 (MJD~50973). It had a maximum count rate of about 82\,cts\,s$^{-1}$\,SSC$^{-1}$, 
which was about 5 times brighter than the count rate seen before and 
after the event (Fig.~\ref{plot_asm}b). It started between UT June 8 22:53:30 and June 9 00:50:40.
The decay is exponential-like with a decay time of $\simeq$1.6\,hr (1.5--12\,keV).
About 4.4~hours after the peak of the flare the count rate (27$\pm$2\,cts\,s$^{-1}$\,SSC$^{-1}$) 
was still above the pre-flare persistent level ($\simeq$17\,cts\,s$^{-1}$\,SSC$^{-1}$), 
suggesting the event lasted even longer. From the end of the last
pre-flare measurement and the first measurement at the persistent count rate after
the flare (June 9 15:05:04) I infer an upper limit on the duration of the
flare of 16.2~hours. 

In Table~1 the fitted exponential decay times of the long flare in the 
total energy band and the three sub-bands are listed.
They were determined using the flare data plus the persistent
data up to a day after the start of the flare. 
The decay time in the 5--12\,keV band is significantly shorter than that
in the 3--5\,keV band, indicating spectral softening during the decay. 
This was verified by determining hardness ratios, using the following procedure.
I first averaged the count rates in the three energy bands of data 
within 30~min bins. Then the ratio of the count rates 
in a hard to low energy band was computed (there are five possible independent 
combinations with the three available energy bands). I then investigated the flare hardness 
values during the decay and compared them with those of the persistent emission. 
It turned out that ratio of the count rates in the 5--12\,keV and 1.5--3\,keV energy bands 
gave the best indication for spectral cooling during the day, see Fig.~\ref{plot_asm}c.
At flare maximum the source is harder with respect to just before the flare; during the 
decay of the flare the source softens.

To estimate the observed maximum flux and the fluence of the flare I performed
X-ray spectral fits to the ASM data (see Sect.~2).
The emission of normal type~I bursts can be well described by emission from a black-body
(Swank et al.\ 1977; Lewin et al.\ 1993). This also holds for the superbursts
(e.g.\ Cornelisse et al.\ 2000). The flare was divided into three intervals which
are indicated by dotted lines in Fig.~\ref{plot_asm}b. 
The average pre-flare persistent emission between MJD\,50971.5 and 50973
was subtracted from the total flare emission per interval. 
A black-body emission model subjected to interstellar absorption was used;
The results of the fits are given in Table~1. 
Since the peak luminosity is well below the Eddington luminosity
(see Sect.~4) the radius of the emission region is expected to be constant during 
the decay. I, therefore, also fitted the spectra from the three intervals 
together, coupling the apparent black-body radius, $R_{\rm bb}$, to one value.
The results of these fits are also given in Table 1; I get $R_{\rm bb}$=5.4$\pm$1.5\,km.
The values of $R_{\rm bb}$ obtained in the above spectral fits are more or less comparable 
to those found during the decay of the radius expansion burst seen from GX\,3+1 
(Kuulkers \&\ van der Klis 2000: $\simeq$7.6\,km at 5\,kpc).
Similar results are derived for e.g.\ KS\,1731$-$260 
(Kuulkers et al.\ 2001).
I, therefore, also performed spectral fits with 
$R_{\rm bb}$ fixed at 7.6\,km, of which the results are also shown in Table 1.
Note that some of the values for $\chi^2_{\rm red}$ are unrealistically low;
this is most likely due to an overestimate of the errors in the data points by including
the rms in the average Crab values.
The spectral fits, however, indicate that the black-body temperature, $kT_{\rm bb}$, 
is in the range 1--2\,keV, and decreases, as expected, during the decay of the flare.

To get the values for the observed maximum flux and fluence, the individual 
dwell net-flare (= total flare minus pre-flare) count rates were converted 
into fluxes, using the count rates 
and average fluxes in the three intervals as obtained in the spectral fits
with $R_{\rm bb}$ left free. The decay in net-flare flux was fitted with an exponential and assumed
it approached zero at infinity. The fluence was calculated by integrating the exponential
from the start of the flare up to infinity. I note that this may underestimate 
the fluence, since the flare probably started somewhat earlier. 
Extrapolating the fitted exponential backwards in time and assuming the 
maximum flux can not exceed the Eddington limit (see Sect.~4) an estimate of the
upper limit on the fluence, $E_{\rm b,max}$, may be obtained. The results are displayed in
Table 1.

\begin{table}
\caption{Flare characteristics}
\begin{tabular}{ccccc}
\hline
\multicolumn{5}{l}{Exponential decay times} \\
\multicolumn{1}{c}{Band} & \multicolumn{1}{c}{$\tau_{\rm exp}$$^a$} & 
\multicolumn{1}{c}{$\chi^2_{\rm red}$/dof} & \multicolumn{2}{c}{~} \\
\multicolumn{1}{c}{(keV)} & \multicolumn{1}{c}{(hr)} & 
\multicolumn{1}{c}{~} & \multicolumn{2}{c}{~} \\
1.5--12 & 1.6 $\pm$ 0.2         & 2.4/21 & & \\
1.5--3  & 1.8 $^{+~1.5}_{-~0.9}$ & 0.9/21 & & \\
3--5    & 2.9 $^{+~1.1}_{-~0.7}$ & 2.3/21 & & \\
5--12   & 1.41 $\pm$ 0.15        & 2.3/21 & & \\
\hline
\multicolumn{5}{l}{X-ray spectral fits: $R_{\rm bb}$ free} \\
\multicolumn{1}{c}{Interval} & \multicolumn{1}{c}{$F_{\rm bb}$$^b$} & 
\multicolumn{1}{c}{$kT_{\rm bb}$ (keV)} & \multicolumn{1}{c}{$R_{\rm bb}$$^c$} &
\multicolumn{1}{c}{$\chi^2_{\rm red}$/dof} \\
1 & 3.0 $\pm$ 0.4 & 2.2 $\pm$ 0.3 & 6 $\pm$ 1      & 1.7/1  \\
2 & 0.9 $\pm$ 0.3 & 2.0 $\pm$ 0.6 & 4 $\pm$ 2      & 0.25/1 \\   
3 & 0.6 $\pm$ 0.1 & 1.2 $\pm$ 0.3 & 8 $^{+~5}_{-~3}$ & 0.01/1 \\
\hline
\multicolumn{5}{l}{X-ray spectral fits: $R_{\rm bb}$ coupled} \\
\multicolumn{1}{c}{Interval} & \multicolumn{1}{c}{$F_{\rm bb}$$^b$} & 
\multicolumn{1}{c}{$kT_{\rm bb}$ (keV)} & \multicolumn{1}{c}{$R_{\rm bb}$$^c$} &
\multicolumn{1}{c}{$\chi^2_{\rm red}$/dof} \\
1 & $\simeq$3.0 & 2.2 $^{+~0.5}_{-~0.3}$ & 5.4 $\pm$ 1.5 & 0.7/5 \\
2 & $\simeq$0.8 & 1.6 $\pm$ 0.3 & 5.4 $\pm$ 1.5 & 0.7/5 \\
3 & $\simeq$0.6 & 1.5 $\pm$ 0.3 & 5.4 $\pm$ 1.5 & 0.7/5 \\
\hline
\multicolumn{5}{l}{X-ray spectral fits: $R_{\rm bb}$ fixed} \\
\multicolumn{1}{c}{Interval} & \multicolumn{1}{c}{$F_{\rm bb}$$^b$} & 
\multicolumn{1}{c}{$kT_{\rm bb}$ (keV)} & \multicolumn{1}{c}{$R_{\rm bb}$$^c$} &
\multicolumn{1}{c}{$\chi^2_{\rm red}$/dof} \\
1 & 2.5 $\pm$ 0.2 & 1.80 $\pm$ 0.04 & 7.6 & 2.5/2 \\
2 & 0.7 $\pm$ 0.1 & 1.31 $\pm$ 0.07 & 7.6 & 1.7/2 \\
3 & 0.6 $\pm$ 0.1 & 1.23 $\pm$ 0.08 & 7.6 & 0.01/2 \\
\hline
\multicolumn{5}{l}{Flare parameters} \\
\multicolumn{1}{c}{$F_{\rm bb,max}$$^d$} & \multicolumn{1}{c}{$E_{\rm b}$$^e$} & 
\multicolumn{1}{c}{$E_{\rm b,max}$$^f$} & 
\multicolumn{1}{c}{$\tau$ (hr)$^g$}  & \multicolumn{1}{c}{~} \\
3.3 $\pm$ 0.6 & $\simeq$1.7 & $\simeq$5.3 & $\simeq$1.5 & \\ 
\hline
\multicolumn{5}{l}{$^a$\,Exponential decay time.} \\
\multicolumn{5}{l}{$^b$\,Unabsorbed black-body flux (10$^{-8}$\,erg\,cm$^{-2}$\,s$^{-1}$).} \\
\multicolumn{5}{l}{$^c$\,Apparent black-body radius at 5\,kpc.} \\
\multicolumn{5}{l}{$^d$\,Peak unabsorbed black-body flux (10$^{-8}$\,erg\,cm$^{-2}$\,s$^{-1}$).} \\
\multicolumn{5}{l}{$^e$\,Fluence in 10$^{-4}$\,erg\,cm$^{-2}$.} \\
\multicolumn{5}{l}{$^f$\,Estimated maximum fluence in 10$^{-4}$\,erg\,cm$^{-2}$, see text.} \\
\multicolumn{5}{l}{$^g$\,$\tau\equiv E_{\rm b}/F_{\rm bb,max}$.} \\
\end{tabular}
\end{table}

An estimate of the recurrence time of such flares is an important constraint on the
physics of the flare. In my case this is very uncertain, given only one observed flare
in 5.8~yrs. One could naively deduce that flares happen every 5.8~yrs, but this obviously 
assumes they occur periodically and that the source was continuously observed in that interval. 
There are, however, about 15600 dwells of 90\,s in that timespan, implying an exposure time of only 16~days;
they are also not equally spaced in time and the number of dwells per day varies wildly up to about 
140 per day. Another flare could thus have 
happened during data gaps of longer than the flare duration, whereas one could rule out
another flare when there are more than one dwell data point with the source at its 
persistent count rate in a time span shorter than the flare duration. I
used the observed flare to estimate the total effective exposure
time during which only one flare occurred, which may be used as a rough indicator on the
recurrence time. The flare profile was characterized
by an exponential in the 1.5--12\,keV band, as determined by the fit to the decay,
with a start point corresponding to
the observed peak of 82\,cts\,s$^{-1}$\,SSC$^{-1}$. A window of 0.2~day was used
(approximately the lower limit on the duration of the flare), which was moved throughout 
all dwell data points. I determined whether or not the flare could fit within the window by 
determining the $\chi^2_{\rm red}$ between the expected profile and the observed data points
(note that I varied the expected persistent count rates and logged the smallest
values of $\chi^2_{\rm red}$).
Whenever $\chi^2_{\rm red}$ exceeded 2.5 (about the same 
value as for the exponential fit to the decay light curve), 
I concluded that within the window no such flare
could have occurred; if it was less than 2.5 a flare could have occurred and only
the individual dwell durations were taken into account. In this way I estimate 
a total effective exposure time of about 490~days. Note that this estimate
is dependent on the used window length. If I use a window of 0.1 or 0.3~day I derive
$\simeq$320~days and $\simeq$600~days, respectively. 

\section{Discussion}

I found an X-ray flare from GX\,3+1 in RXTE/ASM data spanning about 6~yrs; the flare had
a decay time of 1.6\,hr and a duration of 
longer than 4.4\,hr, but shorter than 16.2~hr. 
During the exponential-like decay the flare spectrum softened.
I conclude that the flare has its origin in unstable thermonuclear burning.
The total fluence of the event is between about 
5$\times$10$^{41}$\,erg and 2$\times$10$^{42}$\,erg for a distance of 5\,kpc. 
The maximum net-burst flux reached during a normal type~I X-ray burst seen by
Kuulkers \&\ van der Klis (2000) was 6.9$\times$10$^{-8}$\,erg\,cm$^{-2}$\,s$^{-1}$.
Since this burst was a radius-expansion event the corresponding luminosity at
the neutron star surface reached the Eddington limit. 
The flare, therefore, had an observed maximum of about 0.5 times the Eddington value
(but note that the actual peak of the flare was missed). 
The persistent ASM count rate before and after the flare
was similar to that observed around the radius expansion burst, 
indicating the persistent flux near the X-ray flare was about 0.2 times the Eddington value.

The long X-ray flare bears all the characteristics of the superbursts discovered recently in 
five other X-ray burst sources:
4U\,1735$-$44, Ser\,X-1, 4U\,1636$-$53, 4U\,1820$-$30 and KS\,1731$-$260 
(Cornelisse et al.\ 2000, 2001; Wijnands 2001; Strohmayer \&\ Brown 2001; 
Kuulkers et al.\ 2001). They have been shown to be thermonuclear events on the surfaces of
neutron stars, in which carbon (Cumming \&\ Bildsten 2001;
Strohmayer \&\ Brown 2001) or 
electron capture (Kuulkers et al.\ 2001) 
plays the dominant role. 

Assuming the persistent luminosity is a direct measure of the mass accretion rate, $\dot{M}$, onto
the neutron star, one would infer $\dot{M}$$\simeq$2$\times$10$^{17}$\,g\,s$^{-1}$ in GX\,3+1
just before the superburst. But since the persistent luminosity is variable by a factor of
$\simeq$2 and at the time of the superburst it was at a minimum as seen by the ASM in 
$\simeq$6~yrs, 
I assume an average $\dot{M}$ of $\simeq$3$\times$10$^{17}$\,g\,s$^{-1}$.
If the ignition of the superburst occurs in a carbon-rich layer, recurrence times of 
$\simeq$13~yrs are then expected, unless the ignition is prematurely triggered
(Strohmayer \&\ Brown 2001). Note that this case is applicable to a
neutron star accreting pure helium, such as 4U\,1820$-$30. 
Of the other systems, only 4U\,1735$-$44 and 4U\,1636$-$53 are known to have hydrogen-rich donors
(Augusteijn et al.\ 1998); this is not as yet clear for GX\,3+1.
For hydrogen/helium accreting neutron stars it has been shown that carbon may ignite 
earlier when in an ocean of heavy elements, leading to a smaller recurrence time of $\simeq$2~yrs
at the average $\dot{M}$ (Cumming \&\ Bildsten 2001). 
If, on the other hand, the superbursts are due to unstable electron
capture by protons with subsequent capture of the resulting neutrons by heavy nuclei,
then recurrence times of less than 0.5~yr are expected (Kuulkers et al.\ 2001).
The expected recurrence times in the latter two cases are more or less
compatible with seeing one superburst from GX\,3+1 
in a total effective exposure time of about 1 to 1.5~yrs, 
as well as seeing two such events 4.7~yrs apart from 4U\,1636$-$53 
(Wijnands 2001). Effective exposure times can in principle 
be determined for the other
superburst sources, 
using RXTE/ASM data together with data from other instruments (e.g.\ BeppoSAX/WFCs); however, this is outside the scope of this paper.
Longer and more frequent monitoring of these sources will enable one to discriminate the origin of the superbursts.

\begin{acknowledgements}
I thank Al Levine for providing the ASM on-source effective area matrix and discussions
related to it, Jean in 't Zand for comments on an earlier
draft, Andrew Cumming and Lars Bildsten for a discussion on the recurrence times, and the referee for useful comments.
I acknowledge the use of ASM data products provided by the ASM/RXTE team
({\tt http://xte.mit.edu}; 
{\tt http://heasarc.gsfc.nasa.gov/docs/xte/asm\_products.html}).
\end{acknowledgements}

\end{document}